\documentstyle[pascos,psfig]{article}

\def\be{\begin{equation}}
\def\ee{\end{equation}}
\def\bea{\begin{eqnarray}}
\def\eea{\end{eqnarray}}
\def\[{\left [}
\def\]{\right ]}
\def\({\left (}
\def\){\right )}
\def\lbr{\left\{}
\def\rbr{\right\}}
\def\Tr{{\rm Tr}}
\newcommand{\dilaton}{{\ell}}
\newcommand{\Lag}{{\cal L}}
\newcommand{\superint}{\int \diff^{4}\theta}
\newcommand{\lowest}{|_{\theta =\bar{\theta}=0}}
\newcommand{\diff}{\mbox{d}}

\newcommand{\WaWa}{\Tr({\cal W}^{\alpha}{\cal W}_{\alpha})}

\newcommand{\DbDb}{{\cal D}_{\dot{\alpha}}{\cal D}^{\dot{\alpha}}}

\newcommand{\baal}{b_{a}^{\alpha}}
\newcommand{\baaleff}{\(\baal\)_{\rm eff}}
\newcommand{\bpaleff}{\(b_{+}^{\alpha}\)_{\rm eff}}
\newcommand{\caleff}{\({c_{\alpha}}\)_{\rm eff}}
\newcommand{\lang}{\left\langle}
\newcommand{\rang}{\right\rangle}
\newcommand{\order}{{\cal O}}

\begin{document}

\title{HIDDEN SECTOR GAUGINO
  CONDENSATION AND THE OBSERVABLE WORLD}

\author{BRENT D. NELSON}

\address{Department of Physics, University of California, Berkeley\\
  Ernest Orlando Lawrence Berkeley National Laboratory\\
  Berkeley, California 94720} 


\maketitle\abstracts{We study the phenomenology of a class of models
  describing modular invariant gaugino condensation in the hidden
  sector of a low-energy effective theory derived from the heterotic
  string. Placing simple demands on the resulting observable sector,
  such as a supersymmetry-breaking scale of approximately 1 TeV, results in
  significant restrictions on the possible configurations of the
  hidden sector.}
This talk summarizes recent investigations~\cite{RGEpaper,gaugino} of the
phenomenology of heterotic string-derived supergravity theories
invoking gaugino condensation in a hidden sector to break
supersymmetry and is a condensation of material more fully presented
elsewhere.~\cite{RGEpaper,gaugino,ModInv} The philosophy behind the research
reported here is to ask
whether we can obtain intuition on the nature of the hidden sector matter
and gauge content by requiring ever-increasing
degrees of agreement with the observed world. The effective Lagrangian
we employ incorporates recent developments in 
string theory and tempers a high degree of realism with sufficient
assumptions to preserve tractability. 

Supersymmetry breaking is implemented via condensation of gauginos
charged under the hidden sector gauge group ${\cal G}=\prod_{a}{\cal
  G}_{a}$. For each gaugino
condensate a vector superfield $V_a$ is introduced and the gaugino
condensate superfields $U_{a}\simeq {\WaWa}_{a}$ are then identified as
the (anti-)chiral projections of the vector superfields
$U_{a}=-\(\DbDb-8R\)V_{a}$. The dilaton field (in the linear multiplet 
formalism used here) is the lowest component of the
vector superfield $V={\sum_{a}}V_{a}$: $\dilaton =V{\lowest}$. The
vacuum expectation value (VEV) of this field is related to the unified 
gauge coupling at the string scale by the relation $<\dilaton>=g_{\rm
  str}^{2}/2$. In the class of orbifold compactifications we will be
considering there are, in addition to the dilaton, three untwisted 
moduli chiral superfields $T^{I}$ which parameterize the size of the
compactified space. 

The K\"ahler potential for these moduli and the
matter chiral superfields $\Phi^{A}$ is given by
$K=\ln{V}+\sum_{I}g^{I} +{\sum_A}e^{{\sum_I}{q_{I}^{A}}{g^{I}}} \left| \Phi^{A}
\right|^{2} + \order \( \Phi^{4} \)$,
where $g^{I}=-\ln{(T^{I}+{\overline{T}}^{I})}$ and the $q_{I}^{A}$ are
the modular weights of the fields
$\Phi^{A}$. The relevant part of the complete effective Lagrangian is
then ${\Lag}_{{\rm eff}}={\Lag}_{{\rm KE}} + {\Lag}_{{\rm GS}} +
{\Lag}_{{\rm VY}} + {\Lag}_{{\rm pot}}$.
The kinetic energy Lagrangian
has been ammended to include possible non-perturbative corrections of
string-theoretical origin involving the dilaton: 
${\Lag}_{{\rm KE}}={\superint}E\[-2+f\(V\)\]$. The parameterization of 
these effects that we will adopt is of the the
form~\cite{Shenker,nonpert} $f\(V\)=\[{A_0}+{A_1}/\sqrt{V}\]e^{-B/\sqrt{V}}$,
which was shown~\cite{susybreak} to allow dilaton stabilization at
weak to moderate string coupling with parameters that are all of
$\order \(1\)$.\footnote{As an example, the dilaton potential can be
  minimized with vanishing cosmological constant and $\alpha_{\rm
    str}=0.04$ for $A_{0}=3.25, A_{1}=-1.70$ and $B=0.4$ in $f\(V\)$.}
In the presence of these non-perturbative effects the
relation between 
the string coupling and the VEV of the dilaton becomes ${g_{\rm str}^2}/{2}=
{\dilaton}/{\(1+f\(\dilaton\)\)}$.

The second term in~${\Lag}_{\rm eff}$ is a generalization of the original
Veneziano-Yankielowicz superpotential term for the gaugino condensate,
~\cite{VY}
\begin{equation}
{\Lag}_{\rm VY}=\frac{1}{8}{\sum_a}{\superint}\frac{E}{R}{U_a}
\left [ {b_{a}'} \ln{\(e^{-K/2}{U_a}\)} + {\sum_{\alpha}} {b_{a}^{\alpha}} 
\ln{\[ \( {\Pi^{\alpha}} \)^{p_{\alpha}} \] } \right ] + {\rm h.c.},
\label{eq:LagVY}
\end{equation}
which involves the gauge
condensates $U_a$ as well as possible gauge-invariant matter condensates described by
chiral superfields $\Pi^{\alpha} \simeq {\prod_{A}} {\( \Phi^{A}
  \)}^{n_{\alpha}^{A}}$. The
coeffecients $b_{a}'$, $\baal$ are determined
by demanding the correct transformation properties of the expression
in~(\ref{eq:LagVY}) under chiral and conformal
transformations~\cite{ModInv,match} and yield the following relations:
\begin{eqnarray}
b_{a}\equiv b_{a}' + {\sum_{\alpha}}\baal =\frac{1}{8{\pi}^2}
\(C_{a}-\frac{1}{3} {\sum_A}C_{a}^{A}\),&
{\displaystyle {\sum_{\alpha,A}} {b_{a}^{\alpha}} {n_{\alpha}^{A}}{p_{\alpha}} =
{\sum_{A}}\frac{C_{a}^{A}}{4{\pi}^2}}.
\label{eq:coeff}
\end{eqnarray}
In~(\ref{eq:coeff}) the quantities $C_{a}$ and
$C_{a}^{A}$ are the quadratic Casimir operators for the adjoint and matter representations,
respectively. 

The third term in~${\Lag}_{\rm eff}$ is a modular-invariant
superpotential term for the matter condensates: ${\Lag}_{\rm pot}=
\lbr\frac{1}{2}{\superint}\frac{E}{R}{e^{K/2}}W\[
\( \Pi^{\alpha}\)^{p_{\alpha}} ,T^{I} \] + {\rm h.c.}\rbr$. We adopt
the simplifying assumptions that there are no 
unconfined hidden sector matter fields, each condensate is charged
under only one hidden sector gauge group and that each term in the
hidden sector superpotential is effectively of dimension three. This
allows a simple factorization of the superpotential of the form $W\[
\({\Pi}^{p_{\alpha}}\),T \]={\sum_{\alpha}}c_{\alpha}{W_{\alpha}}\(T\)\(
{\Pi}^{\alpha} \)^{p_{\alpha}}$, and we require ${p_{\alpha}}\sum_{A}
n_{\alpha}^{A}=3 \ \ \ \forall \alpha$.

The remaining term in~${\Lag}_{\rm eff}$ is the
Green-Schwarz (GS) counterterm introduced to ensure modular
invariance. In this note we consider its simplest possible form
${\Lag}_{\rm GS}= b{\superint}EV{\sum_I}g^{I}$
where $b\equiv C_{E_8}/8{\pi}^2 \approx 0.38$ is
proportional to the beta-function coefficient for the group $E_8$.

Having described the effective Lagrangian we are in a position to turn 
our attention to the observable sector phenomenology. The equations of
motion for the auxiliary fields of the condensates $U^{a}$ give
\begin{equation}
{{\rho_a}^2}=e^{-2{\frac{b'_a}{b_a}}}e^{K}e^{-\frac{\(1+f\)}{{b_a}\dilaton}}e^{-\frac{b}{b_a} 
  {\sum_I}g^{I}}{\prod_I}\left|{\eta}\(t^{I}\)\right|^{\frac{4\(b-b_{a}\)}{b_a}}
{\prod_{\alpha}}\left|\baal/4c_{\alpha}\right|^{-2 {\frac{b_{a}^{\alpha}}{b_a}}},
\label{eq:cond1}
\end{equation}
where $t_{I}\equiv T_{I}\lowest$, $u_{a} = U_{a}\lowest \equiv
{\rho}_{a}e^{i{\omega}_a}$ and ${\eta}(t^{I})$ is the Dedekind
function. The scalar potential for the moduli $t_{I}$ is
minimzed at the self-dual points $\lang t_{I} \rang =1$ or $\lang
t_{I} \rang = \exp{\(i\pi/6\)}$, where the corresponding F-components $F_{I}$
of the chiral superfields $T^{I}$ vanish -- thus allowing for the
``dilaton-dominated'' supersymmetry-breaking pattern with naturally
suppressed flavor-changing neutral currents.\cite{dildom} 

To disentangle the complexity of~(\ref{eq:cond1}) it is
convenient to assume that all of the matter in the hidden
sector which transforms under a given subgroup ${\cal G}_a$ is of the
same representation, such as the fundamental representation, and then
make the simultaneous variable redefinition
\begin{eqnarray}
{\sum_{\alpha}}\baal\equiv\baaleff={N_c}b_{a}^{\rm rep}; & \caleff
\equiv {N_c}\({\prod_{\alpha=1}^{N_c}}c_{\alpha}\)^\frac{1}{N_c}.
\label{eq:baaleff}
\end{eqnarray}
In the above equation $b_{a}^{\rm rep}$ is proportional to the
quadratic Casimir operator for the matter fields in the common
representation and $N_c$ is the number of condensates.

From a determination of the condensate value $\rho$
the supersymmetry-breaking scale can be found
by solving for the gravitino mass, given by
$M_{3/2}=\frac{1}{3}\lang\left|M\right|\rang=\frac{1}{4}\lang
\left|{\sum_a}{b_a}{u_a} \right|\rang$. In the case of multiple gaugino
condensates the scale of supersymmetry breaking is governed by the
condensate with the largest one-loop beta-function
coefficient.~\cite{RGEpaper}
Hence we will here consider the case with just one
condensate with beta-function coefficient denoted $b_+$:
$M_{3/2}=\frac{1}{4}{b_+}\lang\left|u_+\right|\rang$.

\begin{figure}[b,t,h]
\centerline{
       \psfig{file=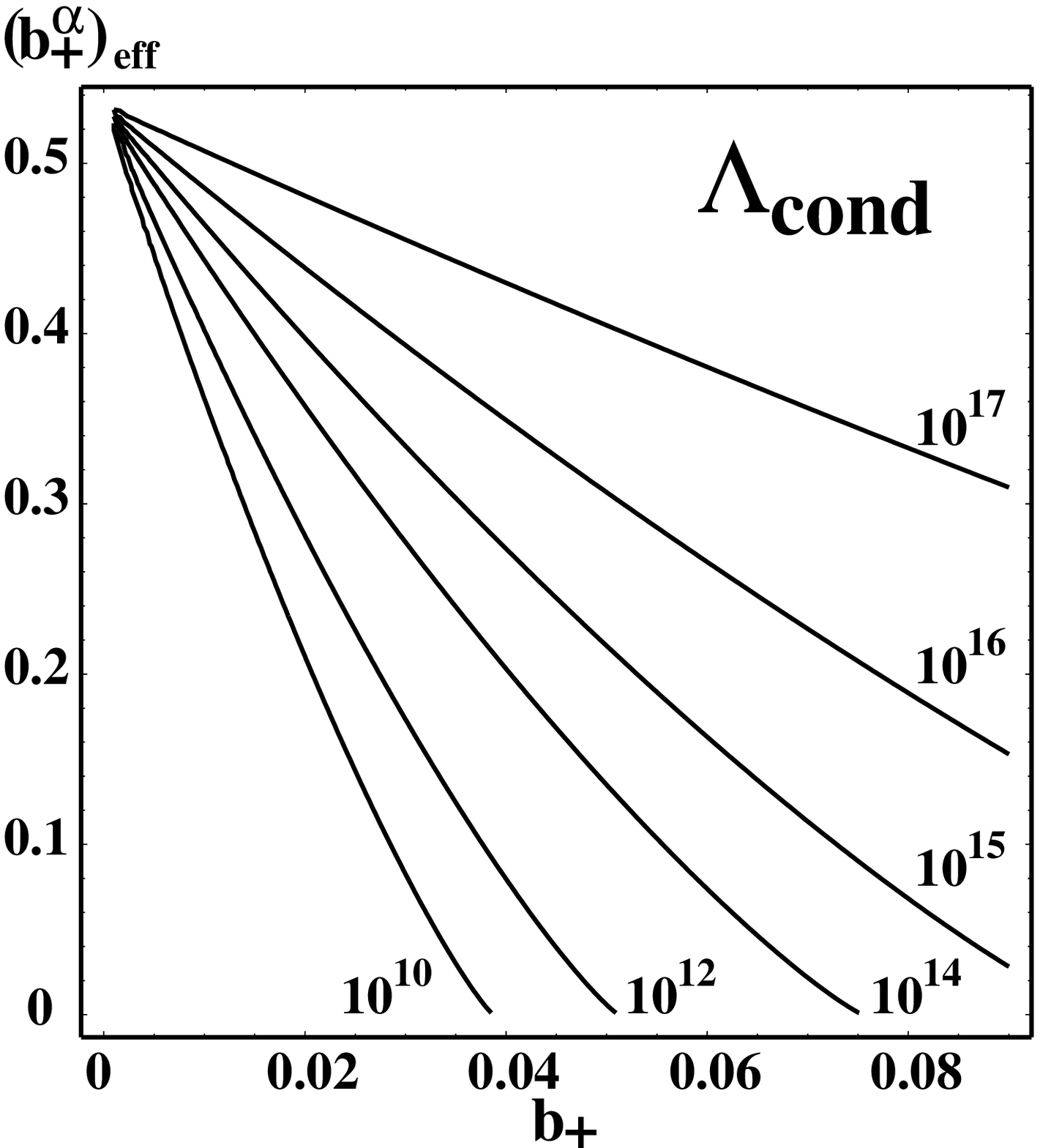,width=0.4\textwidth}
       \psfig{file=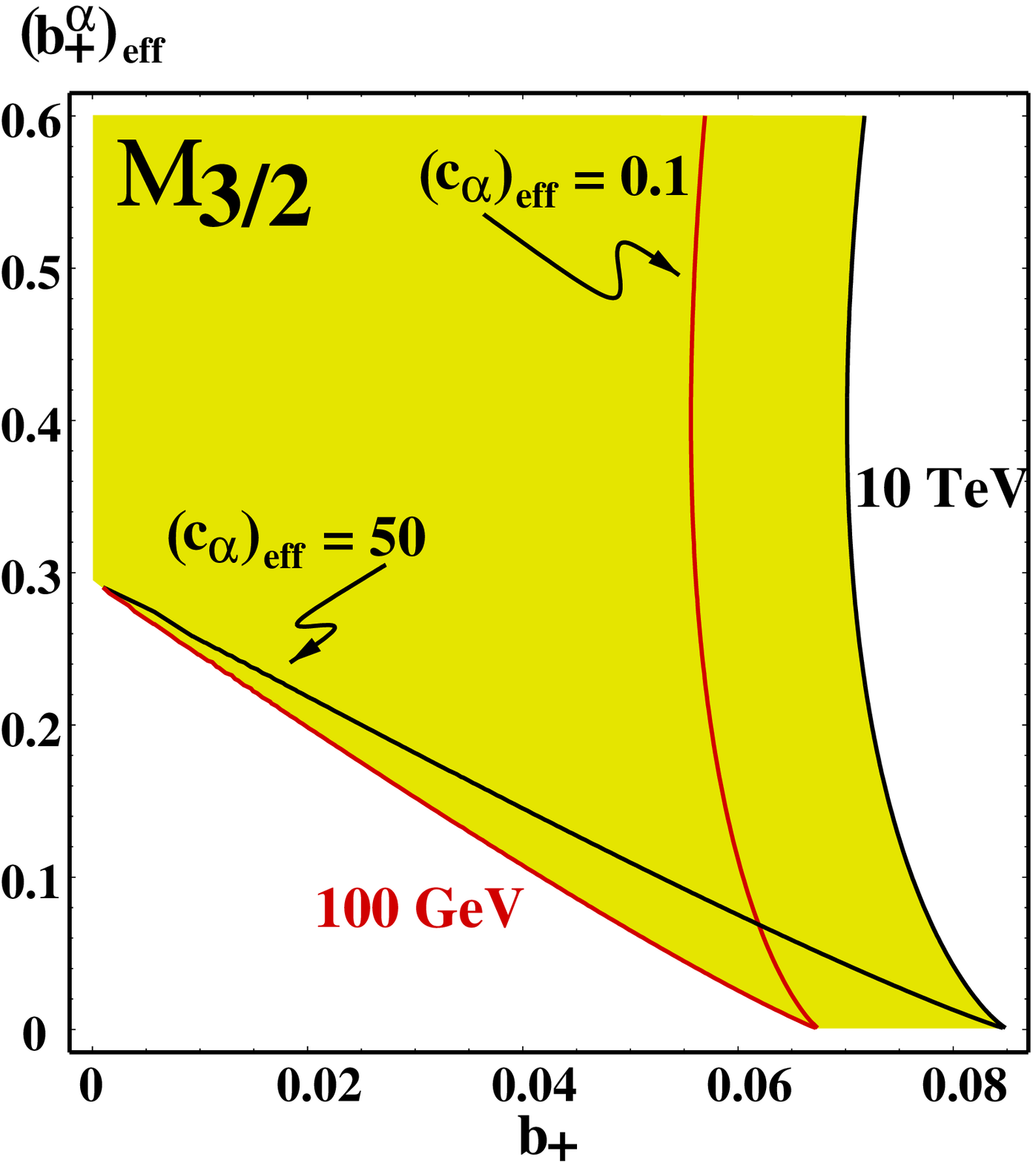,width=0.4\textwidth}}
          \caption{{\footnotesize {\bf Gaugino Condensation and
                Gravitino Mass}. The sharp
variation of the condensation scale (in GeV) with the parameters of
the theory, is apparent in the contour plot on the left. The right
panel demonstrates 
the dependence of the supersymmetry-breaking scale on the Yukawa
couplings of the hidden sector.}}
        \label{fig:cond}
\end{figure}

Now for given values of $\caleff$ and $g_{\rm str}$ the condensation
scale $\Lambda_{\rm cond}=\(M_{\rm Planck}\)\lang {\rho}^{2}_{+}\rang^{1/6}$
and gravitino mass can be plotted in the $\lbr b_{+},
\bpaleff\rbr$ plane, as in Figure~\ref{fig:cond}. Also shown is the
variation of the gravitino mass as a function
of the Yukawa parameters $\caleff$, where a very generous spread in 
the supersymmetry-breaking scale and hidden sector Yukawas is allowed.

\begin{figure}[b,t,h]
\centerline{
       \psfig{file=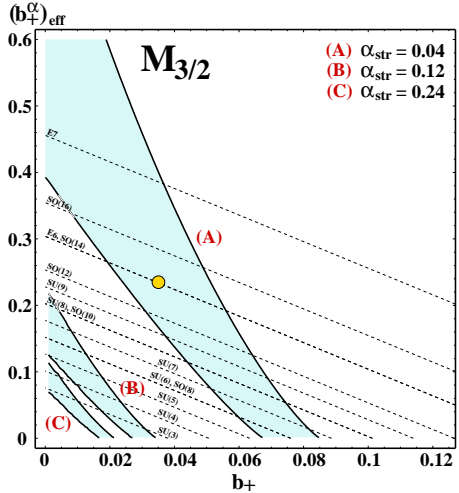,width=0.5\textwidth}}
          \caption{{\footnotesize {\bf Constraints on the Hidden
                Sector}. The shaded regions give three different
              ``viable'' regions depending on the value of the unified
              coupling strength at the string scale. The upper limit
              in each case represents a 10~TeV gravitino mass contour
              with $\caleff=1$, while the lower bound represents a
              100~GeV gravitino mass contour with $\caleff=10$. The
              highlighted point is an example of a hidden sector with an $E_6$ subgroup and 9
              {\bf 27}s of matter.}}
        \label{fig:bestplot}
\end{figure}

Upon $Z_N$ orbifold compactification the $E_8$ gauge group of the hidden
sector is presumed to break to some subgroup(s) of $E_8$. For each
such subgroup the equations in~(\ref{eq:baaleff}) define a 
line in the $\lbr b_{+},\bpaleff\rbr$ plane. In
Figure~\ref{fig:bestplot} we have overlaid these gauge lines on a
plot similar to the previous one. We restrict the effective Yukawa 
couplings of the hidden sector to a more reasonable range and give
three different values of the string
coupling at the string scale. The choice of string coupling is made
when specifying the
boundary conditions for solving the dilaton potential, and
hence $f(\lang\dilaton\rang)$ in equation~(\ref{eq:cond1}).

A typical matter configuration would be represented in
Figure~\ref{fig:bestplot} by a point on one of the gauge group
lines. The number of possible configurations consistent with a given choice of
$\lbr \alpha_{\rm str}, \caleff \rbr$ and supersymmetry-breaking scale 
$M_{3/2}$ is quite restricted. For example, 
Figure~\ref{fig:bestplot} immediately rules out hidden sector gauge
groups smaller
than SU(6) for weak coupling at the string scale $\(g^{2}_{\rm str} \simeq 
0.5\)$. Furthermore, even moderately larger values of the string
coupling at unification become difficult to obtain as it
is necessary to postulate a hidden sector with very small gauge group
and particular combinations of matter to force the beta-function
coefficient to small values.

\begin{figure}[t]
\centerline{
       \psfig{file=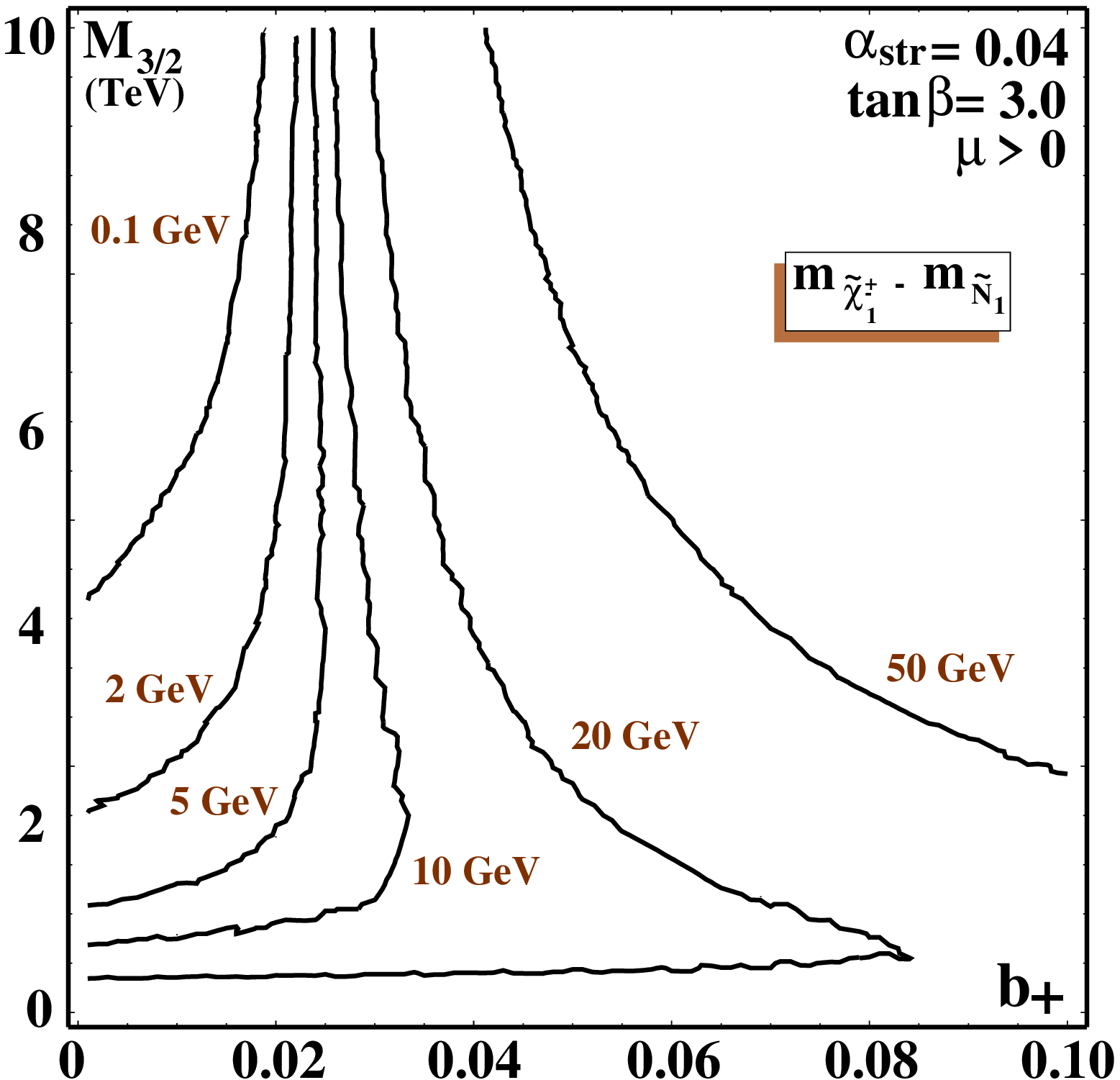,width=0.5\textwidth}
       \psfig{file=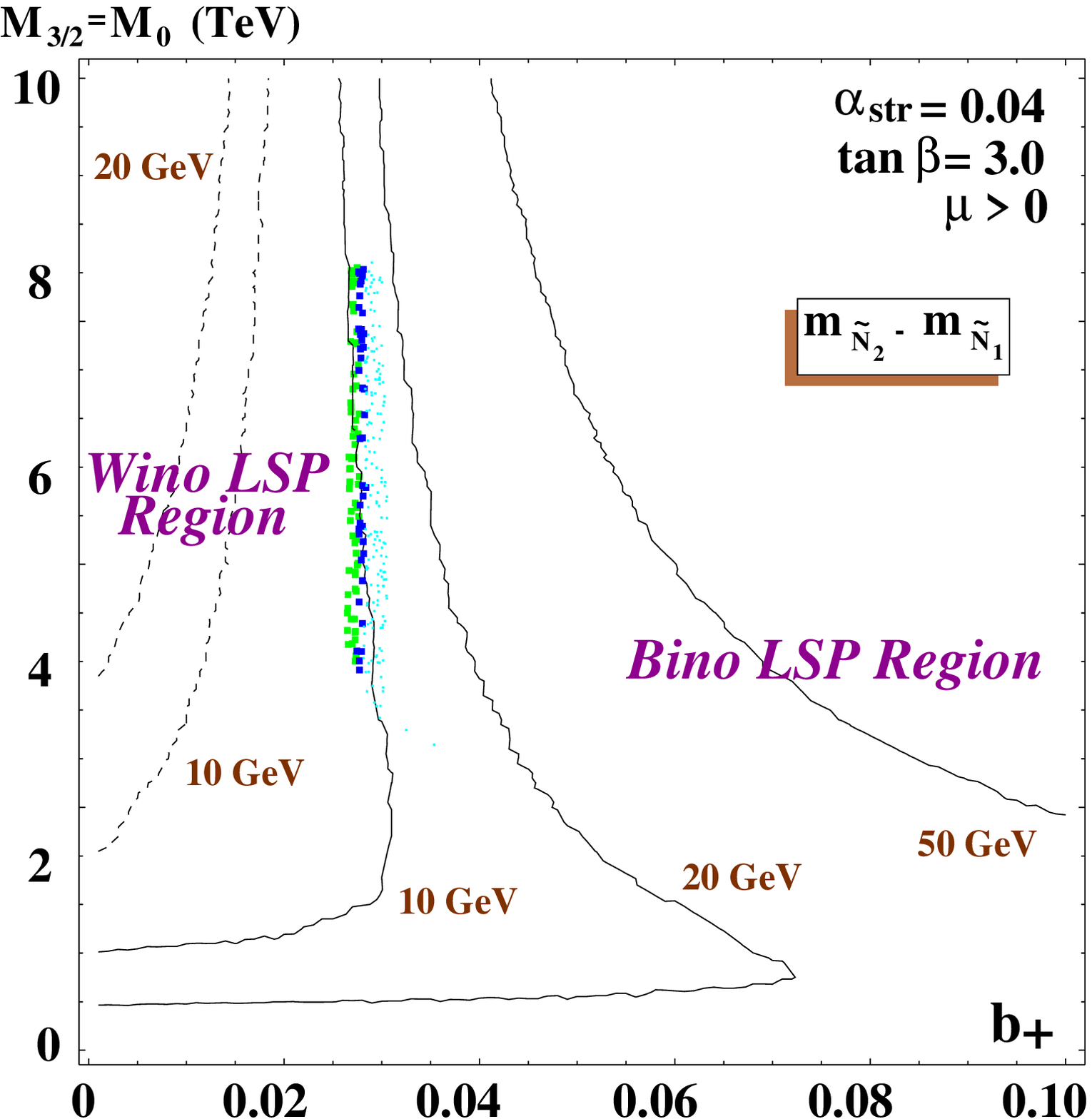,width=0.5\textwidth}}
          \caption{{\footnotesize {\bf The Physical Gaugino
                Sector}. The left panel gives the mass difference
              between the $\chi^{\pm}_{1}$ and the $N_1$ in
              GeV. The right
              panel gives the difference in mass between the two
              lightest neutralinos $N_{2}$ and $N_{1}$. Note that a
              level crossing occurs and there exists a region in which 
              the $W_{0}$ becomes the LSP, as is to be expected when
              the anomaly contribution to gaugino masses
              dominates. The scatter points represent parameter space
              for which the relic density lies in the range $0.1 \leq
              \Omega_{\chi} h^2 \leq 1$.}}
        \label{fig:gauginosector}
\end{figure}

Additional constraints on the model arise from demanding an acceptable
pattern of electroweak symmetry breaking, super-partner
masses consistent with search results from LEP and the Tevatron and
cosmology. The spectra of soft-terms arising from these models was
investigated at tree-level~\cite{RGEpaper} and recently at the
one-loop level.~\cite{masspaper}  To focus on the gaugino
sector, the gaugino masses at the condensation scale are given by 
\begin{equation}
m_{{\lambda}_a}|_{\mu = \Lambda_{\rm cond}} =
-\frac{g^{2}_{a}\(\mu\)}{2} \[\frac{3{b_+}\(1+{b'_a}\ell\)}{1+{b_+}\ell}
-3{b_a}\] M_{3/2},
\label{gauginomass}
\end{equation}
when the matter fields do not couple to the Green-Schwarz
counterterm. The first term in~(\ref{gauginomass}) arises from the
presence of the gaugino condensate directly, while the second term is
the one-loop contribution arising from the conformal anomaly. When the
condensing group beta-function coefficient $b_{+}$ is sufficiently
small this anomaly-induced piece can be the dominant contribution to
gaugino masses, giving rise to a phenomenology typical of so-called
``anomaly-mediated'' models.~\cite{anomaly} Such situations generally
produce a neutralino as the lightest supersymmetric particle (LSP)
with significant wino-like content, as is
displayed in Figure~\ref{fig:gauginosector}.

In models where the scalars decouple from the Green-Schwarz term the
scalar masses are equal to the gravitino mass and typically are larger than the
gaugino masses by an order of magnitude. Thus, avoiding LEP search
limits on charginos generally requires scalar masses in the TeV
range and low $\tan\beta$.~\cite{RGEpaper} In standard supergravity
scenarios such heavy scalars generally
prove catastrophic as relic neutralino LSPs fail to annihilate rapidly
enough in the early universe to be consistent with current knowledge
of the age of the universe ({\it i.e.} $\Omega_{\chi}h^2 \leq
1$).~\cite{EFO} However, if the LSP has a small but significant
wino-like content this problem can be avoided and the relic LSP can
again be an excellent candidate for cold dark matter. In the right
panel of Figure~\ref{fig:gauginosector} we have overlaid those points
in parameter space for which the relic density of neutralinos is in
the cosmologically allowed range of $0.1 \leq \Omega_{\chi} h^2 \leq 1$.
Note that the scalars can be multi-Tev in mass provided the condensing
group beta-function coefficient lies in a particular range --
suggestively the range in which the highlighted point of
Figure~\ref{fig:bestplot} lies. In this cosmologically preferred
region of the parameter space the wino content of the LSP is
10-20\%. Additional analysis of models with heavy scalars and
satisfactory relic densities is underway.~\cite{andreas}

A more realistic
model may alter these results to some degree and uncertainty remains in 
the general size and nature of the Yukawa couplings of the hidden
sector of these theories. Nevertheless this survey suggests that
eventual measurement of the size and pattern of supersymmetry breaking 
in our observable world may well imply a very limited choice
of hidden sector configurations (and 
hence string compactifications) compatible with low energy
phenomena.

\section*{Acknowledgments}
This work was performed in collaboration with Mary K. Gaillard and was 
supported in part by the U.S. Department of Energy under 
Contract DE-AC03-76SF00098 and in part by the National Science 
Foundation under grant PHY-95-14797 and PHY-94-04057.

\end{document}